\documentclass[aps,reprint,amsmath,amssymb,superscriptaddress,nofootinbib,floatfix]{revtex4-2}

\usepackage{bm}
\usepackage{mathtools}
\usepackage{booktabs}
\usepackage{xcolor}
\usepackage{tikz}
\usetikzlibrary{arrows.meta,positioning,calc,fit,shapes.geometric,decorations.pathmorphing}
\usepackage{hyperref}
\hypersetup{hidelinks}

\newcommand{\Hh}{\mathcal H}
\newcommand{\I}{\mathbb I}
\newcommand{\R}{\mathbb R}
\newcommand{\ket}[1]{\left|#1\right\rangle}
\newcommand{\bra}[1]{\left\langle#1\right|}
\newcommand{\braket}[2]{\left\langle#1\middle|#2\right\rangle}

\newcommand{\Tr}{\operatorname{Tr}}

\newcommand{\teachingbox}[1]{%
\begin{center}
\fbox{\begin{minipage}{0.92\columnwidth}\small\textbf{Teaching checkpoint.} #1\end{minipage}}
\end{center}}

\begin{document}

\title{What do position and time mean in the quantum wavefunction?}

\author{Mustafa Bakr}
\affiliation{Clarendon Laboratory, Department of Physics, University of Oxford, Oxford OX1 3PU, United Kingdom}
\author{Zichi Zhang}
%\affiliation{Clarendon Laboratory, Department of Physics, University of Oxford, Oxford OX1 3PU, United Kingdom}
\author{Margot Stakenborg}
%\affiliation{Clarendon Laboratory, Department of Physics, University of Oxford, Oxford OX1 3PU, United Kingdom}

%\date{\today}

\begin{abstract}
The notation $\psi(x,t)$ is among the first pieces of quantum mechanics that students learn. It is also among the easiest to over-interpret. Because $x$ and $t$ occur as arguments of the same function, students may ask whether they have the same mathematical status. They may also ask whether $\psi(t)$ should require a generalized bra $\bra{t}$ in the same way that $\psi(x)=\braket{x}{\psi}$ is often written. A related question is whether the absence of a universal time operator follows simply from Pauli's argument. These questions mix several structures that are usually introduced in different parts of the curriculum. We present a unified pedagogical treatment built around two maps hidden in $\psi(x,t)$. Time evolution selects a state along a trajectory in Hilbert space. A spectral representation then maps that state to amplitudes labelled by outcomes of a chosen observable. We formulate the position representation without generalized eigenkets. We recover Dirac's $\ket{x}$ notation as a controlled continuum shorthand and use a finite-grid limit to show where delta normalization enters. We distinguish background coordinates, translation parameters, observables, spectral labels, and physical records. We also clarify the Stone-theorem analogy, compare prescribed-time position measurements with arrival-time measurements, state what the strong form of Pauli's argument excludes, and exhibit an exactly solvable boundary case in which a canonical self-adjoint time observable exists. Spin, circuit-QED, and optical-clock examples provide experimentally grounded checks. The aim is not a new interpretation of time. It is a reusable teaching framework for separating mathematical role from notation.
\end{abstract}

\maketitle

\section{Introduction}

A student encountering the position wavefunction is taught to write
\begin{equation}
\psi(x)=\braket{x}{\psi}.
\label{eq:position_amp_intro}
\end{equation}
Soon afterward the same student writes $\psi(x,t)$, or even informally speaks of ``$\psi(t)$.'' This raises a natural question. If $x$ enters through a bra $\bra{x}$, should $t$ enter through a bra $\bra{t}$? If not, why do the two symbols sit side by side as arguments of the same function?

The question is elementary to state, but it sits at the intersection of several topics that are often taught separately. These include abstract states and representations, the spectral theorem and continuous spectra, Stone's theorem, time evolution, quantum clocks, generalized measurements, and the limitations of Pauli's argument~\cite{Pauli1933}. Physics-education research has documented student difficulties in translating between Dirac states and wavefunction representations~\cite{MarshmanSingh2015}. Emigh, Passante, and Shaffer also documented difficulties in applying time dependence to quantum systems and in recognizing the role of the energy eigenbasis~\cite{Emigh2015}. The conjunction of these difficulties motivates the present treatment.

We do not claim that position representations, time observables, or the problem of time are new subjects. The relativity of quantum representations has long been emphasized~\cite{delaTorre2002}. Equivalent formulations of quantum mechanics have also been systematically compared by Styer \emph{et al.}~\cite{Styer2002}. Hilgevoord has also stressed that the slogan ``position is an operator while time is a parameter'' can compare unlike objects. Background spacetime coordinates and dynamical observables must first be distinguished~\cite{Hilgevoord2002,Hilgevoord2005}. The temporal column of our taxonomy follows Busch's threefold distinction between external, dynamical, and event time~\cite{Busch1990,Busch2008}. This is its closest ancestor. Quantum clocks and event-time observables likewise have an extensive literature~\cite{Peres1980,MugaLeavens2000,MugaBook2008}. Our narrower pedagogical goal is to expose the different maps that ordinary notation compresses into $\psi(x,t)$. We then turn that distinction into a sequence of classroom examples.

The central message can be stated in one sentence.
\begin{quote}
\emph{Two symbols can be arguments of the same written function without entering the theory through the same mathematical or physical operation.}
\end{quote}

For a time-independent Hamiltonian, the construction is
\begin{equation}
\ket{\psi_0}
\xrightarrow{\,U(t)\,}
\ket{\psi_t}
\xrightarrow{\,W_X\,}
\psi_t(x).
\label{eq:two_maps}
\end{equation}
Here $U(t)$ moves along a dynamical trajectory in Hilbert space, whereas $W_X$ chooses the spectral representation of the position observable. Equation~\eqref{eq:two_maps}, rather than the superficial symmetry of the notation $\psi(x,t)$, will organize the paper.

\begin{figure*}[t]
\centering
\begin{tikzpicture}[>=Latex, font=\small, x=1cm,y=1cm]
  % Hilbert-space region
  \draw[rounded corners, thick] (-5.7,0.3) rectangle (5.7,3.7);
  \node[anchor=north west] at (-5.45,3.45) {abstract Hilbert space $\Hh$};
  \draw[thick,->] (-4.5,1.5) .. controls (-2.4,3.15) and (0.1,0.85) .. (4.4,2.55);
  \fill (-3.5,2.05) circle (2.2pt) node[above left] {$\ket{\psi_{t_1}}$};
  \fill (0.0,2.05) circle (2.2pt) node[above] {$\ket{\psi_{t_2}}$};
  \fill (3.55,2.45) circle (2.2pt) node[above right] {$\ket{\psi_{t_3}}$};

  % arrows to reps
  \draw[->,thick] (-3.5,1.82) -- (-3.5,-0.25) node[midway,left] {$W_X$};
  \draw[->,thick] (0,1.82) -- (0,-0.25) node[midway,left] {$W_X$};
  \draw[->,thick] (3.55,2.04) -- (3.55,-0.25) node[midway,right] {$W_X$};

  % wave functions
  \foreach \xc/\ph in {-3.5/0,0/35,3.55/70}{
     \draw[->] (\xc-1.35,-1.6) -- (\xc+1.35,-1.6) node[right] {$x$};
     \draw[->] (\xc-1.15,-2.35) -- (\xc-1.15,-0.55);
     \draw[thick,domain=-1.05:1.05,samples=60,smooth]
       plot ({\xc+\x},{-1.58+0.62*exp(-1.8*(\x-0.18*sin(\ph))*(\x-0.18*sin(\ph)))*cos(2.7*\x+\ph)});
  }
  \node at (0,-2.75) {$x$ labels amplitudes in the chosen position representation};
\end{tikzpicture}
\caption{The two maps hidden in $\psi(x,t)$. Time evolution selects a member $\ket{\psi_t}$ of a family of abstract states. Only afterward does a representation map $W_X$ turn that state into a function of the spectral coordinate $x$. The same notation $\psi(x,t)$ compresses both operations.}
\label{fig:two_maps}
\end{figure*}
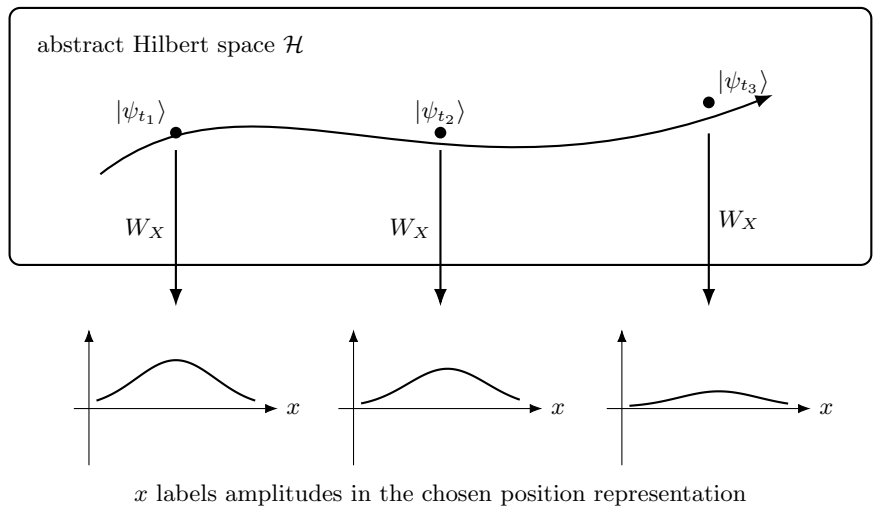

\section{The two maps hidden in $\psi(x,t)$}
\label{sec:twomaps}

Let $H$ be a time-independent self-adjoint Hamiltonian. Stone's theorem associates it with a strongly continuous unitary group
\begin{equation}
U(t)=e^{-iHt/\hbar}.
\end{equation}
Given an initial state $\ket{\psi_0}$, the state at the parameter value $t$ is
\begin{equation}
\ket{\psi_t}=U(t)\ket{\psi_0}.
\label{eq:evolution}
\end{equation}
No representation has yet been chosen. Equation~\eqref{eq:evolution} makes sense for a spin, a harmonic oscillator, a finite-dimensional qudit, or a particle on a line.

Now choose a position observable $X$. In the familiar multiplicity-one case, the spectral theorem gives a unitary representation map
\begin{equation}
W_X:\Hh\longrightarrow L^2(\R,dx)
\end{equation}
for which $X$ acts as multiplication by $x$. The position-space wavefunction is therefore
\begin{equation}
\psi_t(x)=\left(W_X\ket{\psi_t}\right)(x)
           =\left(W_XU(t)\ket{\psi_0}\right)(x).
\label{eq:composition}
\end{equation}
The notation $\psi(x,t)$ is a convenient rewriting of $\psi_t(x)$. It should not be read as evidence that $x$ and $t$ entered through identical constructions.

This distinction remains if the Hamiltonian depends explicitly on time. One then replaces the one-parameter group by a propagator $U(t,t_0)$, but $t$ still indexes the dynamical family while $x$ indexes the chosen representation.

\teachingbox{Ask students to remove position entirely. A spin-$1/2$ state $\ket{\psi(t)}=a(t)\ket{\uparrow}+b(t)\ket{\downarrow}$ still evolves with $t$. The role of Schr\"odinger time therefore cannot originate from the distributional subtleties of the continuous kets $\ket{x}$.}

\section{Position representation without generalized kets}
\label{sec:no_kets}

A next natural question is whether $\ket{x}$ is essential. It is not. The spectral measure of $X$ is enough.

For a self-adjoint position operator $X$, let $E_X(\Delta)$ denote the projection associated with a measurable spatial set $\Delta$. The spectral theorem states
\begin{equation}
X=\int_{\R} x\,dE_X(x),
\label{eq:spectraltheorem}
\end{equation}
and the probability of finding a normalized state in $\Delta$ is
\begin{equation}
\Pr(X\in\Delta)=\bra{\psi}E_X(\Delta)\ket{\psi}.
\label{eq:pvmprob}
\end{equation}
Equations~\eqref{eq:spectraltheorem} and \eqref{eq:pvmprob} require no vector $\ket{x}$ belonging to $\Hh$.

In the standard position representation, the projection simply multiplies a wavefunction by the indicator function of the region,
\begin{equation}
(W_XE_X(\Delta)\psi)(x)=\mathbf 1_\Delta(x)\psi(x).
\end{equation}
The familiar Born density then follows when the spectral measure is absolutely continuous.
\begin{equation}
\Pr(X\in\Delta)=\int_\Delta |\psi(x)|^2dx.
\label{eq:bornx}
\end{equation}
Thus one can teach the position representation as a unitary spectral representation first and introduce $\ket{x}$ only later as compact Dirac notation. Rigged-Hilbert-space treatments make the generalized-ket formalism precise~\cite{Roberts1966}, but they are not logically required for an introductory statement of the spectral theorem.

There is a subtle point worth making explicit. A generic element of $L^2(\R)$ is an equivalence class of functions that agree almost everywhere. Point evaluation $\psi\mapsto\psi(x_0)$ is therefore not a bounded functional on all of $L^2$. Writing $\braket{x_0}{\psi}$ as if $\bra{x_0}$ were an ordinary Hilbert-space bra silently assumes extra structure. This is one reason delta-normalized eigenkets are generalized objects rather than ordinary normalizable states. Gieres~\cite{Gieres2000} and de la Madrid~\cite{delaMadrid2005} survey these pitfalls of the bra--ket formalism, and the rigged-Hilbert-space structures that resolve them, at a level suitable for instructors.

\subsection{A finite-grid route to the Dirac delta}

A useful classroom bridge avoids distributions at first and lets them emerge from an ordinary finite-dimensional basis. The continuum completeness relation and delta normalization approached by this construction are standard~\cite{Shankar1994}; here we introduce an explicit finite-grid rescaling so that the normalization bookkeeping is visible. Divide a line into grid cells of width $\Delta x$ and let $\ket{j}$ be orthonormal states,
\begin{equation}
\braket{j}{k}=\delta_{jk},\qquad
\sum_j\ket{j}\bra{j}=\I.
\end{equation}
Define continuum-scaled labels
\begin{equation}
\ket{x_j}_{c}\equiv\frac{\ket{j}}{\sqrt{\Delta x}}.
\end{equation}
Then
\begin{equation}
{}_{c}\!\braket{x_j}{x_k}_{c}=\frac{\delta_{jk}}{\Delta x},
\qquad
\sum_j\Delta x\,\ket{x_j}_{c}{}_{c}\!\bra{x_j}=\I.
\label{eq:grididentity}
\end{equation}
For a state $\ket{\psi}=\sum_j c_j\ket{j}$, define
\begin{equation}
\psi(x_j)={}_{c}\!\braket{x_j}{\psi}=\frac{c_j}{\sqrt{\Delta x}}.
\end{equation}
Normalization becomes
\begin{equation}
\sum_j\Delta x\,|\psi(x_j)|^2=1.
\end{equation}
In the continuum limit, the three structures approach
\begin{align}
\frac{\delta_{jk}}{\Delta x}&\longrightarrow\delta(x-x'),\\
\sum_j\Delta x&\longrightarrow\int dx,\\
\sum_j\Delta x\ket{x_j}\bra{x_j}&\longrightarrow
\int dx\,\ket{x}\bra{x}=\I,
\end{align}
where the last expressions are understood weakly or distributionally.

This exercise answers a common question precisely. One can avoid generalized vectors by staying with the spectral measure or a finite-grid approximation. Once one manipulates a point-labelled $\ket{x}$ with delta normalization, however, the distributional structure has been reintroduced implicitly.

It also prevents a notational mistake. The standard identity is not $\int\ket{x}\,dx$. Rather,
\begin{equation}
\I=\int dx\,\ket{x}\bra{x},
\qquad
\ket{\psi}=\int dx\,\ket{x}\psi(x),
\label{eq:diracweak}
\end{equation}
with the appropriate generalized/weak interpretation.

\teachingbox{Give students Eq.~\eqref{eq:grididentity} before mentioning rigged Hilbert spaces. Ask which power of $\Delta x$ is required so that probabilities remain finite as the grid is refined. The Dirac delta then appears as a normalization necessity rather than a mysterious axiom.}

% Original heading (typo, empty section):
% \section{the relaionships between different roles of x/t}
\section{The relationships between the different roles of $x$ and $t$}
\label{sec:roadmap}

Table~\ref{tab:taxonomy} in Sec.~\ref{sec:taxonomy} separates five roles that the letters $x$ and $t$ each play. Five mathematical connections are important for understanding the differences. We treat them in the following sections.

\begin{enumerate}
\item \emph{The spectral theorem} ties an observable to its spectral/outcome labels and to its representation label. For position, the operator $X$, the outcome $x$, and the argument of $\psi(x)$ form one unit (Secs.~\ref{sec:twomaps} and \ref{sec:no_kets}).
\item \emph{Covariance} ties an observable to the corresponding translation parameter. Equation~\eqref{eq:covariance} relates $X$ to the displacement $a$ (Sec.~\ref{sec:stone}).
\item \emph{Stone's theorem} ties a translation parameter to its generator. The pairings are $a\leftrightarrow P$ and $\tau\leftrightarrow H$ in Eq.~\eqref{eq:stoneanalogy} (Sec.~\ref{sec:stone}).
\item \emph{Operational calibration} ties the temporal translation parameter to measurements. The Schr\"odinger $t$ is identified with elapsed time because physical clocks are dynamical systems that order evolution consistently (Secs.~\ref{sec:stone} and \ref{sec:symmetric}).
\item \emph{Naimark dilation} ties an event-time observable to a time representation label. A covariant time POVM supplies a time-labelled amplitude function through an isometry into a larger space (Sec.~\ref{sec:symmetric}).
\end{enumerate}

The asymmetry of the standard formalism can now be stated. For position, link~1 is available in unitary form. For the semibounded Hamiltonians emphasized below, the corresponding time package is generally available through link~5 in dilated form rather than as a unitary spectral representation on the original Hilbert space. Conflations arise when one of these links is replaced by an identity. One example is identifying the spectral label $x$ with the translation parameter $a$. Another is identifying the group parameter $\tau$ with a clock reading.

\section{Which $x$ and which $t$?}
\label{sec:taxonomy}

The phrase ``space versus time'' can conceal several comparisons. Hilgevoord emphasized that background coordinates and dynamical observables should not be conflated~\cite{Hilgevoord2002,Hilgevoord2005}. Busch~\cite{Busch1990,Busch2008} sharpened the temporal side of this comparison into a threefold distinction between external time, dynamical time, and event time. The temporal column of Table~\ref{tab:taxonomy} refines that classification. Table~\ref{tab:taxonomy} separates five roles that are frequently denoted by the same letters.

\begin{table*}[t]
\caption{A taxonomy of labels that are easily conflated. }
\label{tab:taxonomy}
\small
\setlength{\tabcolsep}{2pt}
\begin{tabular}{p{2.9cm}p{4.85cm}p{4.85cm}p{3.35cm}}
\toprule
Role & Spatial example & Temporal example & Typical mathematical object\\
\midrule
Background coordinate & Coordinate $x$ used to describe fields and apparatus & Coordinate $t$ used to order  & Parameter\\
Translation parameter & Displacement $a$ in a spatial translation & Time displacement $\tau$ in a stationary dynamical translation & One-parameter group label\\
System observable & Particle position (operator) $X$ & Clock reading, arrival time, dwell time, or another event-time observable & PVM or POVM, with a self-adjoint operator in the sharp case\\
Spectral/outcome label & Possible outcome $x$ of measuring $X$ & Possible reading $\tau$ of a clock/time POVM & Point in outcome space\\
Representation label & $x$ in the $X$-spectral representation $\psi(x)$ & A label $\tau$ only if a time observable/clock representation has actually been chosen & Coordinate on a representation/outcome space\\
\bottomrule
\end{tabular}
\end{table*}

This taxonomy explains why the slogan ``position is an operator, time is a parameter'' is both useful and dangerous. In ordinary nonrelativistic quantum mechanics, the position of the system is indeed usually represented as an observable while an external clock parameter labels the Schr\"odinger evolution. But background spatial coordinates are parameters too, and a physical clock reading can itself become a quantum observable. The apparent asymmetry depends on which rows of Table~\ref{tab:taxonomy} are being compared.

A further distinction is useful here. For position, the spectral theorem connects the observable $X$ with its spectral labels $x$ and with the corresponding $X$-representation. It does not identify the spectral label $x$ with the translation parameter $a$. The latter belongs to a different structure, the unitary translation group. It is related to $X$ by covariance. Thus $X$, $x$, and $a$ are closely related but are not the same mathematical object. For time, the Schr\"odinger label $t$ is already available as an evolution parameter without first introducing a time observable. Only when a physical clock, arrival-time measurement, or other time observable is specified does a separate spectral/outcome label arise. This is why one should not infer a universal $\bra{t}$ merely from the notation $\psi(x,t)$.

The same caution applies already in classical Hamiltonian mechanics. Ordinary time is normally an external evolution parameter, not a phase-space coordinate canonically conjugate to the Hamiltonian. One can enlarge the phase space and promote time to a coordinate with a conjugate momentum, but that is an extended construction rather than the starting definition of classical time~\cite{Belot2007}. This is why the question ``Why is time not simply the canonical coordinate of $H$?'' should not begin from the assumption that classical mechanics already treats it that way.

\section{Stone's theorem and the tempting analogy}
\label{sec:stone}

Stone's theorem gives a particularly clean diagnostic. Spatial translations form a one-parameter unitary group
\begin{equation}
T(a)=e^{-iaP/\hbar},
\end{equation}
whose generator is momentum $P$. Temporal translations for a time-independent system form
\begin{equation}
U(\tau)=e^{-i\tau H/\hbar},
\end{equation}
whose generator is the Hamiltonian $H$.

The direct analogy is therefore
\begin{equation}
\boxed{\begin{aligned}
\text{spatial displacement }a &\leftrightarrow P,\\
\text{time displacement }\tau &\leftrightarrow H
\end{aligned}}
\label{eq:stoneanalogy}
\end{equation}
It is not $X\leftrightarrow t$.

The position observable is related to the translation group by covariance. With a conventional sign choice,
\begin{equation}
T(a)^\dagger X T(a)=X+a\I.
\label{eq:covariance}
\end{equation}
Equation~\eqref{eq:covariance} says that translating the physical state relative to the coordinate frame changes the position observable's statistics in the expected way. It does not identify $X$ with the group parameter $a$.

Stone's theorem by itself also does not explain why its group parameter should be called physical time. The theorem supplies a mathematical parameter and generator. Physics identifies that parameter with elapsed laboratory time because the same parameter consistently orders the evolution of physical systems and is calibrated operationally by clocks. A clock is itself a dynamical system.

\teachingbox{Ask, ``What is the spatial analogue of Schr\"odinger time?'' The answer is not the position operator. It is the parameter of spatial translation. This single question often reveals whether a student is mixing a group parameter with an observable.}

\subsection{A useful cross-check from field theory in lower spatial dimension}

There is a useful way to run the logic in the opposite direction. In relativistic field theory one normally labels a field by spacetime coordinates. For example, one writes $\phi(t,x,y,z)$ in $3+1$ dimensions. Reducing the number of spatial dimensions gives $\phi(t,x,y)$ in $2+1$, then $\phi(t,x)$ in $1+1$, and finally a single dynamical coordinate $q(t)$ in $0+1$ dimensions. The last case is ordinary quantum mechanics written in path-integral language. There is no spatial field coordinate left, but the history is still labelled by time.

This is standard in field-theory pedagogy. In the Oxford C6 notes, Chalker and Lukas describe the functional-integral coordinate as one-dimensional and representing time in quantum mechanics. In quantum field theory it becomes a spacetime vector~\cite{ChalkerLukas2010}. The same point opens Zee's textbook development of the path integral~\cite{Zee2010}. Their formulation is a useful consistency check on the distinction developed here. The presence of $t$ as a label of histories does not require a time operator or a spectral time basis.

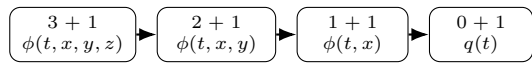
\begin{figure}[t]
\centering
\begin{tikzpicture}[>=Latex,font=\scriptsize,node distance=2.7mm]
\node[draw,rounded corners,align=center,text width=1.45cm] (a) {$3+1$\\$\phi(t,x,y,z)$};
\node[draw,rounded corners,right=of a,align=center,text width=1.35cm] (b) {$2+1$\\$\phi(t,x,y)$};
\node[draw,rounded corners,right=of b,align=center,text width=1.25cm] (c) {$1+1$\\$\phi(t,x)$};
\node[draw,rounded corners,right=of c,align=center,text width=1.15cm] (d) {$0+1$\\$q(t)$};
\draw[->,thick] (a)--(b);\draw[->,thick] (b)--(c);\draw[->,thick] (c)--(d);
\node[align=center,below=4mm of c] {remove spatial coordinates, not the evolution label};
\end{tikzpicture}
\caption{A dimensional route from field theory back to quantum mechanics. In the standard notation, ``$0+1$ dimensional'' means zero spatial dimensions plus one time dimension. A genuinely zero-dimensional field theory, with no time coordinate either, is instead an ordinary finite-dimensional integral or matrix-model-type object, not quantum mechanics.}
\label{fig:dimensionalroute}
\end{figure}

This argument should not be read as a proof that time can never be an observable. It says something narrower. Ordinary quantum mechanics can be viewed as the zero-spatial-dimensional member of the same history-based formalism. The variable $t$ remains a coordinate on the histories even when no spatial representation label is present. A separate clock or event-time measurement can still introduce time-valued outcomes, as discussed below.

The same route also runs forward. It shows how relativistic field theory achieves a more even-handed treatment of $x$ and $t$ by demoting $x$. In quantum field theory the dynamical object is the field $\phi(x,t)$, which is an operator-valued distribution. Both arguments are background coordinates in the sense of the first row of Table~\ref{tab:taxonomy}. They label operators rather than spectral outcomes of a particle position observable. No covariant particle-position observable of the ordinary nonrelativistic kind is generally available. The Newton--Wigner construction~\cite{NewtonWigner1949} does supply a self-adjoint position operator for a single relativistic particle, but its localization notion is frame-dependent. Hegerfeldt-type results show that, under positive-energy assumptions, states initially localized in a bounded region generally develop instantaneous spatial tails~\cite{Hegerfeldt1974}. Malament's theorem gives a distinct no-go result for a relativistic quantum mechanics of localizable particles under its locality assumptions~\cite{Malament1996}. These localization obstructions are not a spatial version of Pauli's semibounded-spectrum argument: a spatial momentum component $P_i$ is not semibounded merely because the energy--momentum spectrum lies in the forward cone. At the level of local detection events, however, space and time can be treated more even-handedly: a detector event records a where and a when as outcomes. This is Protocol~B of Fig.~\ref{fig:arrival} rather than Protocol~A.

\section{Three physical examples}

\subsection{A spin as a worked example of time without a time basis}
\label{subsec:spin}

A two-level spin gives the cleanest possible example because it removes every complication associated with a continuous spectrum. Consider a spin-$1/2$ in a static magnetic field along $z$, with
\begin{equation}
H=\frac{\hbar\omega}{2}\sigma_z,
\qquad
\ket{\psi(0)}=\frac{\ket{+z}+\ket{-z}}{\sqrt2}.
\end{equation}
Time evolution gives
\begin{equation}
\ket{\psi(t)}=\frac{e^{-i\omega t/2}\ket{+z}+e^{+i\omega t/2}\ket{-z}}{\sqrt2}.
\label{eq:spinworked}
\end{equation}
Nothing in Eq.~\eqref{eq:spinworked} requires a ``time basis.'' The parameter $t$ labels which abstract state on the orbit generated by $H$ is under discussion. If we choose the $\sigma_z$ representation, the spectral labels are simply the two possible outcomes $+1$ and $-1$,
\begin{equation}
\psi_t(+)=\braket{+z}{\psi(t)},\qquad
\psi_t(-)=\braket{-z}{\psi(t)}.
\end{equation}
The state therefore has a time label even though its representation has only two discrete coordinates.

The geometry makes the distinction especially transparent. The initial state points along $+x$ on the Bloch sphere. Evolution under $H\propto\sigma_z$ rotates that Bloch vector around the $z$ axis. Time tells us where the state is along this orbit. It does not tell us which result a subsequent measurement must return. If we measure $\sigma_x$ at laboratory time $t_0$, the possible recorded outcomes remain $+1$ and $-1$, while their probabilities depend on where the state has reached. The probabilities are
\begin{equation}
P(+x,t_0)=\cos^2(\omega t_0/2),\qquad
P(-x,t_0)=\sin^2(\omega t_0/2).
\label{eq:spinprob}
\end{equation}
Thus $t_0$ is prescribed by the experimentalist, whereas $\pm1$ are random spectral outcomes of the chosen observable.

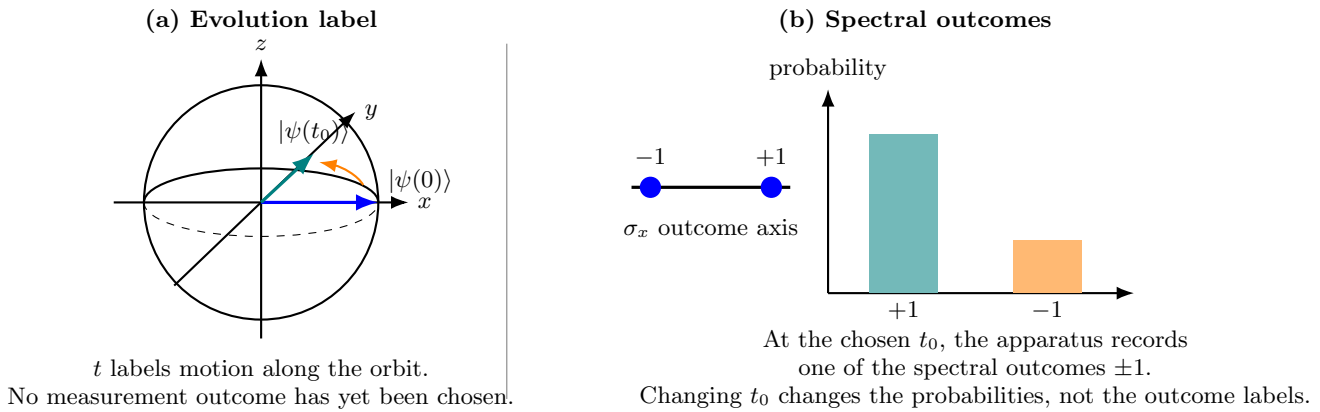
\begin{figure*}[t]
\centering
\begin{tikzpicture}[>=Latex,font=\small,x=1cm,y=1cm]
% Panel A
\node[font=\bfseries\small] at (-4.8,3.75) {(a) Evolution label};
\draw[thick] (-4.8,1.35) circle (1.55);
\draw[->,thick] (-4.8,-0.45)--(-4.8,3.25) node[above] {$z$};
\draw[->,thick] (-6.75,1.35)--(-2.85,1.35) node[right] {$x$};
\draw[->,thick] (-5.95,0.25)--(-3.55,2.55) node[right] {$y$};
\draw[dashed] (-6.35,1.35) arc[start angle=180,end angle=360,x radius=1.55,y radius=0.45];
\draw[thick] (-3.25,1.35) arc[start angle=0,end angle=180,x radius=1.55,y radius=0.45];
\coordinate (O) at (-4.8,1.35);
\coordinate (V0) at (-3.25,1.35);
\coordinate (V1) at (-4.1,2.0);
\draw[->,very thick,blue] (O)--(V0) node[above right,text=black] {$\ket{\psi(0)}$};
\draw[->,very thick,teal] (O)--(V1) node[above,text=black] {$\ket{\psi(t_0)}$};
\draw[->,orange,thick] (-3.45,1.58) arc[start angle=12,end angle=62,x radius=1.15,y radius=0.45];
\node[align=center] at (-4.8,-1.05) {$t$ labels motion along the orbit.\\No measurement outcome has yet been chosen.};

% divider
\draw[gray] (-1.55,-1.25)--(-1.55,3.45);

% Panel B
\node[font=\bfseries\small] at (3.85,3.75) {(b) Spectral outcomes};
\draw[very thick] (0.1,1.55)--(2.2,1.55);
\fill[blue] (0.35,1.55) circle (4pt);\node[above] at (0.35,1.72) {$-1$};
\fill[blue] (1.95,1.55) circle (4pt);\node[above] at (1.95,1.72) {$+1$};
\node[below] at (1.15,1.25) {$\sigma_x$ outcome axis};
\draw[->,thick] (2.7,0.15)--(2.7,2.85) node[above] {probability};
\draw[->,thick] (2.7,0.15)--(6.75,0.15);
\fill[teal!55] (3.25,0.15) rectangle (4.15,2.25);
\fill[orange!55] (5.15,0.15) rectangle (6.05,0.85);
\node at (3.7,-0.08) {$+1$};
\node at (5.6,-0.08) {$-1$};
\node[align=center] at (4.65,-0.85) {At the chosen $t_0$, the apparatus records\\one of the spectral outcomes $\pm1$.\\Changing $t_0$ changes the probabilities, not the outcome labels.};
\end{tikzpicture}
\caption{A two-level control example that separates the roles without any continuous-spectrum subtleties. (a) Schr\"odinger time labels the position of the abstract state along its unitary orbit. (b) Once an observable is selected, its spectrum supplies the possible measurement labels. The laboratory chooses $t_0$. The apparatus returns a random spectral outcome.}
\label{fig:spinroles}
\end{figure*}

\teachingbox{Pause at Fig.~\ref{fig:spinroles} before discussing position. Ask the class, ``At $t_0$, what number is actually written into the data file?'' The answer to a $\sigma_x$ measurement is $+1$ or $-1$, not $t_0$. Repeating the experiment at different $t_0$ maps the time dependence of the probabilities.}

\subsection{A translated wavepacket and three different spatial quantities}

Suppose a wavepacket has a probability density centered near $x_0$. Apply a translation by the amount $a$. The parameter $a$ labels a unitary transformation, the operator $P$ generates it, and the observable $X$ describes where the particle may be found. These are three objects even though all are connected to ``space.'' After translation, the measured position distribution shifts, but no student should call the translation parameter itself the position operator.

\subsection{Position at a chosen time versus time at a chosen position}

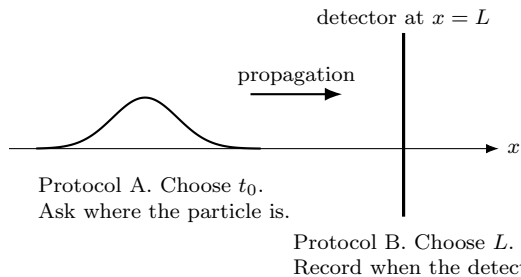
\begin{figure}[b]
\centering
\begin{tikzpicture}[>=Latex,font=\footnotesize,scale=0.9]
  \draw[->] (-0.2,0) -- (7,0) node[right] {$x$};
  \draw[very thick] (5.6,-1.0) -- (5.6,1.7);
  \node[above] at (5.6,1.7) {detector at $x=L$};
  \draw[thick,domain=0.2:3.5,samples=60,smooth]
   plot (\x,{0.75*exp(-2.2*(\x-1.8)*(\x-1.8))});
  \draw[->,thick] (3.35,0.8) -- (4.7,0.8) node[midway,above] {propagation};
  \node[align=left,anchor=west] at (0.1,-0.75) {Protocol A. Choose $t_0$.\\Ask where the particle is.};
  \node[align=left,anchor=west] at (3.85,-1.55) {Protocol B. Choose $L$.\\Record when the detector clicks.};
\end{tikzpicture}
\caption{Two experiments that use the word ``time'' differently. In protocol A, $t_0$ is externally prescribed and position is the random outcome. In protocol B, the detector location is prescribed and the registered event time is the random outcome. The density $|\psi(x,t)|^2$ directly answers the first question, not in general the second.}
\label{fig:arrival}
\end{figure}

The distinction becomes especially physical in a detector experiment. Consider a wavepacket approaching a screen at $x=L$, as sketched in Fig.~\ref{fig:arrival}.

Protocol A asks, ``If I measure position at the externally chosen instant $t_0$, where is the particle?'' The standard answer is the density $|\psi(x,t_0)|^2$.

Protocol B asks, ``The detector is fixed at $L$. When will it click?'' The time of the event is now the random measurement outcome. An arrival-time distribution is a different quantum-measurement problem. In general, it is not obtained by simply reinterpreting $|\psi(L,t)|^2$ as a normalized probability density in $t$. Allcock analyzed the obstructions facing an ideal arrival-time measurement~\cite{Allcock1969}. Kijowski developed an axiomatic arrival-time distribution~\cite{Kijowski1974}. The review of Muga and Leavens~\cite{MugaLeavens2000} surveys the subsequent literature.

The ambiguity is physical enough to support experimental proposals. Bohmian mechanics defines arrival times directly through trajectories. Das and D\"urr computed spin-dependent Bohmian arrival-time distributions and proposed measuring them~\cite{DasDurr2019}. Goldstein, Tumulka, and Zangh\`{\i} subsequently argued that these particular distributions are not generated by a POVM and therefore cannot be the measurement statistics of the proposed quantum measurement~\cite{GoldsteinTumulkaZanghi2024}. Either way, the episode underlines the lesson of this subsection. The time-of-arrival observable is fixed by the measurement model.

The distinction is pedagogically useful because the same laboratory setup can exchange which quantity is prescribed and which quantity is recorded.

% Original headings (lowercase):
% \section{the symmetric construction}
% \subsection{can time be a spectral label?}
\section{The symmetric construction}
\label{sec:symmetric}

%\subsection{Can position be a label of a state?}

\subsection{Can time be a spectral label?}

By symmetry, one is tempted to write down the mirror image of Eq.~\eqref{eq:composition},
\begin{equation}
\psi_a(t)=\bigl(W_T\ket{\psi_a}\bigr)(t)=\bigl(W_T\,T(a)\ket{\psi_0}\bigr)(t),
\label{eq:mirror}
\end{equation}
in which the roles of the two maps are exchanged. A spatial displacement selects the state, and a time representation supplies the amplitude labels. We make two remarks about Eq.~\eqref{eq:mirror}.

First, the notation must respect Table~\ref{tab:taxonomy}. The state family is generated by the spatial translation group $T(a)$ of Eq.~\eqref{eq:stoneanalogy}. The subscript is therefore the group parameter $a$, not the spectral label $x$. Writing $U(x)$ would identify a group parameter with an outcome label. It would also overload the symbol reserved for time evolution. With that bookkeeping in place, the family $\ket{\psi_a}=T(a)\ket{\psi_0}$ exists for every state, exactly as $\ket{\psi_t}$ does.

Secondly, for the semibounded Hamiltonians at issue here, $W_T$ cannot in general be supplied as a unitary spectral representation in which the dynamics acts by translation of $t$. Such a unitary map to $L^2(\R,dt)$ would require a self-adjoint operator $T$ globally conjugate to $H$. Section~\ref{sec:pauli} explains the resulting obstruction. A generalized construction remains available. A covariant time POVM $F_T$ admits a Naimark dilation. In the multiplicity-one case, the dilation can be represented by an isometry $W_T\colon\Hh\to L^2(\R,dt)$ that intertwines the dynamics with translations of $t$; more generally an auxiliary multiplicity space may be required~\cite{Holevo1982,Busch2008}. Kijowski's free-particle construction provides a concrete passage-time observable~\cite{Kijowski1974}. For a semibounded $H$, the mirror construction is therefore naturally realized in dilated rather than globally unitary form. In this construction the residual space--time asymmetry appears in the fact that $W_T$ maps onto a proper subspace rather than unitarily onto the whole representation space. It is not an absence of every possible time-labelled amplitude function. Section~\ref{subsec:clockexample} exhibits $F_T$ and $W_T$ in closed form for a two-level system.

\subsection{Can time be an observable?}

Yes, provided the word time now refers to a physical reading or event variable rather than merely to the external parameter of Eq.~\eqref{eq:evolution}. A general quantum measurement with time-valued outcomes may be described by a POVM $F_T$ satisfying
\begin{equation}
F_T(\Delta)\ge0,\qquad F_T(\R)=\I,
\end{equation}
and
\begin{equation}
\Pr(T\in\Delta)=\bra{\psi}F_T(\Delta)\ket{\psi}.
\label{eq:timepovm}
\end{equation}
A projection-valued measure is the special sharp case associated with a self-adjoint operator. Generalized measurements therefore make clear why ``observable'' need not mean ``self-adjoint operator with a projection-valued measure (PVM)'' in every experimental context.

A physical clock makes the distinction even sharper. If the clock is kept outside the quantum description, its reading supplies the external $t$ used to label $\ket{\psi_t}$. If the clock is included as a quantum subsystem, its pointer can have uncertainty and can become entangled with the system. A conditional or relational description may then ask for the state of the system given a clock reading. This is the relational construction of Page and Wootters~\cite{PageWootters1983}, analyzed and extended by Giovannetti, Lloyd, and Maccone~\cite{GLM2015}. This does not make the external Schr\"odinger parameter and the clock observable mathematically identical. It changes which degrees of freedom are included in the quantum model.

Consequently, saying that ``a state has a definite time but not a definite position'' is potentially misleading. The ordinary state is labelled by the chosen external time parameter. It does not possess that $t$ as an eigenvalue in the same sense in which a position eigenstate may possess a sharp value of $X$.

\subsection{A worked example of the two-level clock}
\label{subsec:clockexample}

The following is a worked example built on Sec.~\ref{subsec:spin}. Let $H=\tfrac{\hbar\omega}{2}\sigma_z$, so the physical dynamics is periodic up to an overall phase and a time reading can be treated as a cyclic variable on $[0,2\pi/\omega)$. Define the time states
\begin{equation}
\ket{\tau}\equiv e^{-i\omega\tau/2}\ket{+z}+e^{+i\omega\tau/2}\ket{-z},
\label{eq:timestates}
\end{equation}
which satisfy $U(s)\ket{\tau}=\ket{\tau+s}$ but are not orthogonal, $\braket{\tau}{\tau'}=2\cos[\omega(\tau-\tau')/2]$. The kets themselves are anti-periodic over one cycle, $\ket{\tau+2\pi/\omega}=-\ket{\tau}$, while the effects $\ket{\tau}\bra{\tau}$ and hence all clock probabilities are periodic. The canonical covariant time POVM of this clock is
\begin{equation}
F_T(\Delta)=\int_\Delta \frac{\omega}{2\pi}\,\ket{\tau}\bra{\tau}\,d\tau .
\label{eq:clockpovm}
\end{equation}
Two one-line integrals confirm $F_T\bigl([0,2\pi/\omega)\bigr)=\I$ and the covariance $U(s)^\dagger F_T(\Delta)\,U(s)=F_T(\Delta-s)$, with the outcome set understood modulo one period. For generic nontrivial $\Delta$, the overlapping time states give effects that are not projections. Equation~\eqref{eq:clockpovm} is therefore a POVM rather than a PVM~\cite{Holevo1982,Busch2008}. In finite dimensions the trace of a commutator vanishes, while $\Tr(i\hbar\I)=2i\hbar$. Hence $[T,H]=i\hbar\I$ has no solution on $\mathbb{C}^2$.

The dilation of Eq.~\eqref{eq:clockpovm} exhibits the $W_T$ of Eq.~\eqref{eq:mirror}. The map $(W_T\psi)(\tau)=\sqrt{\omega/2\pi}\,\braket{\tau}{\psi}$ sends the two-dimensional Hilbert space isometrically onto the subspace of $L^2\bigl([0,2\pi/\omega)\bigr)$ spanned by $e^{\pm i\omega\tau/2}$, and satisfies $(W_T U(s)\psi)(\tau)=(W_T\psi)(\tau-s)$, with the same anti-periodic continuation understood when the translated argument crosses the chosen interval boundary. Evolution becomes translation of $\tau$, exactly as position becomes multiplication under $W_X$, while the range of $W_T$ remains only two-dimensional rather than all of $L^2$.

Finally, for the precessing state $\ket{\psi(t_0)}$ of Eq.~\eqref{eq:spinworked}, the reading is distributed as
\begin{equation}
p(\tau)=\frac{\omega}{\pi}\cos^{2}\!\left[\frac{\omega(\tau-t_0)}{2}\right],
\label{eq:clockreading}
\end{equation}
normalized on one period, peaked at the external time $t_0$, and spread over a sizable fraction of the period. A two-level clock therefore tracks laboratory time with limited resolution. The slogan ``a clock reading is an observable while Schr\"odinger $t$ is a parameter'' becomes quantitative. The parameter $t_0$ enters the distribution of the observable's outcomes as its location.

\section{What Pauli's argument actually excludes}
\label{sec:pauli}

The common textbook slogan that ``Pauli proved there is no time operator''~\cite{Pauli1933} is too strong. The sharp obstruction concerns a stronger set of assumptions. Suppose there were a self-adjoint operator $T$ whose exponentials generated arbitrary energy translations in the Weyl sense. Schematically, one would have
\begin{equation}
e^{isT/\hbar}He^{-isT/\hbar}=H-s\I
\label{eq:pauli_shift}
\end{equation}
for every real $s$ (the sign depends on convention). Equation~\eqref{eq:pauli_shift} translates the spectrum of $H$ by an arbitrary real amount. A Hamiltonian with a lowest energy cannot have a spectrum invariant under every such translation. Thus a globally covariant self-adjoint canonical time operator is incompatible with a semibounded Hamiltonian.

The physically important word is ``globally.'' A formal commutator
\begin{equation}
[T,H]=i\hbar\I
\end{equation}
on a restricted domain does not automatically exponentiate to Eq.~\eqref{eq:pauli_shift} for all real parameters. Galapon and others have emphasized these domain distinctions and constructed time operators under weaker conditions~\cite{Galapon2002}. Time POVMs and operational arrival-time observables also remain possible~\cite{MugaLeavens2000,MugaBook2008}.

Therefore the difference between a semibounded Hamiltonian and momentum unbounded in both directions is relevant to the strong canonical time-operator obstruction. It is not the sole explanation for why standard nonrelativistic quantum mechanics initially uses an external $t$ to label evolution.

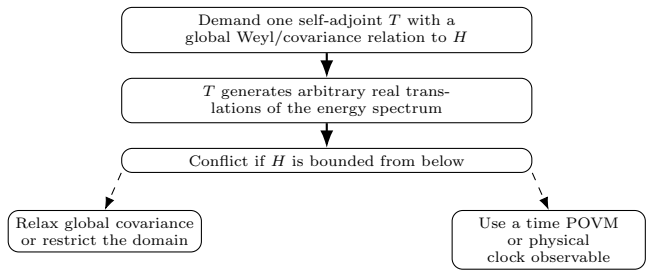
\begin{figure}[t]
\centering
\begin{tikzpicture}[>=Latex,font=\scriptsize,node distance=4mm,scale=0.82,transform shape]
\node[draw,rounded corners,align=center,text width=6.4cm] (ass)
 {Demand one self-adjoint $T$ with a global Weyl/covariance relation to $H$};
\node[draw,rounded corners,below=of ass,align=center,text width=6.4cm] (shift)
 {$T$ generates arbitrary real translations of the energy spectrum};
\node[draw,rounded corners,below=of shift,align=center,text width=6.4cm] (conf)
 {Conflict if $H$ is bounded from below};
\draw[->,thick] (ass)--(shift);
\draw[->,thick] (shift)--(conf);
\node[draw,rounded corners,below left=6mm and -13mm of conf,align=center,text width=2.9cm] (r1)
 {Relax global covariance or restrict the domain};
\node[draw,rounded corners,below right=6mm and -13mm of conf,align=center,text width=2.9cm] (r2)
 {Use a time POVM or physical clock observable};
\draw[->,dashed] (conf.south west)--(r1.north);
\draw[->,dashed] (conf.south east)--(r2.north);
\end{tikzpicture}
\caption{An assumption map for Pauli's argument. The obstruction is not a blanket prohibition on every object called a time observable. It follows from demanding a strong global canonical relation together with a semibounded Hamiltonian.}
\label{fig:pauli}
\end{figure}

\subsection{A solvable boundary case: when a canonical time observable exists}
\label{subsec:canonicaltime}

It is useful to complement the obstruction above with a model in which the obstruction is absent. Consider the ideal translation Hamiltonian
\begin{equation}
H=vP,\qquad v\neq0,
\label{eq:linearH}
\end{equation}
on $L^2(\R)$, where $P$ is the usual self-adjoint momentum operator. The model may be viewed as an ideal one-way or chiral translation model. It should not be confused with a positive-energy massless-particle Hamiltonian $H=c|P|$. Here $\sigma(P)=\R$, so $\sigma(H)=\R$ and the semiboundedness assumption behind the spectral contradiction in Eq.~\eqref{eq:pauli_shift} is absent.

Define
\begin{equation}
\Theta\equiv\frac{X}{v}.
\label{eq:canonicaltime}
\end{equation}
Since $X$ and $P$ are the usual canonical pair on the line,
\begin{equation}
[\Theta,H]
=\left[\frac{X}{v},vP\right]
=i\hbar\I.
\label{eq:canonicaltimecomm}
\end{equation}
Moreover, this is not merely a formal restricted-domain commutator. The Weyl relation of the full-line $X$--$P$ pair gives
\begin{equation}
e^{is\Theta/\hbar}He^{-is\Theta/\hbar}=H-s\I,
\label{eq:canonicaltimeshift}
\end{equation}
with no spectral contradiction because $\sigma(H)=\R$ is invariant under arbitrary real shifts.

The same structure is transparent in the dynamics. The time-evolution operator is
\begin{equation}
U(t)=e^{-iHt/\hbar}=e^{-ivtP/\hbar}=T(vt),
\end{equation}
so that $U(t)\ket{x}=\ket{x+vt}$ with the translation convention of Eq.~\eqref{eq:covariance}. Equivalently,
\begin{equation}
U(t)^\dagger\Theta U(t)=\Theta+t\I.
\label{eq:canonicaltimecovariance}
\end{equation}
In the position representation the Schr\"odinger equation becomes the transport equation
\begin{equation}
\frac{\partial\psi}{\partial t}
+v\frac{\partial\psi}{\partial x}=0,
\end{equation}
with the shape-preserving solution
\begin{equation}
\psi(x,t)=\psi_0(x-vt).
\label{eq:transportsolution}
\end{equation}

The example must still be interpreted using the distinctions of Table~\ref{tab:taxonomy}. The external parameter $t$ continues to label the unitary family $U(t)$; it has not itself become an operator. The self-adjoint observable $\Theta=X/v$ instead has spectral outcomes $\tau=x/v$. What is special in this model is that the observable and the external evolution parameter are connected by the exact covariance relation in Eq.~\eqref{eq:canonicaltimecovariance}. Thus the absence of a globally canonical self-adjoint time observable is not a logical necessity of Hilbert-space quantum mechanics. It is the combination of the desired global canonical relation with spectral properties such as semiboundedness that produces the Pauli obstruction. If Eq.~\eqref{eq:linearH} is regarded as a linearized effective chiral model, its spectrum being unbounded below is correspondingly an idealization rather than a claim about the full spectrum of a stable physical system.

\teachingbox{Ask students to identify the two quantities called time in this example. The external $t$ labels the evolution $U(t)$, while a measurement of $\Theta$ returns a random outcome $\tau$. Even when a self-adjoint canonical time observable exists, covariance connects these two roles rather than making them the same mathematical object.}

\section{Laboratory measurement examples in circuit QED and an optical atomic clock}
\label{sec:labexamples}

\subsection{Circuit QED with a quantum state, classical record, and readout path}
\label{subsec:cqed}

The same distinction between a label in a mathematical representation and a physically registered outcome appears in contemporary experiments. In dispersive circuit-QED readout, a qubit is coupled to a microwave resonator~\cite{Blais2004,Blais2021}. Different qubit states produce different linear-response boundary loads and hence different resonator responses. An outgoing microwave field becomes a state-dependent pointer that can be amplified and recorded. A boundary-condition formulation makes this conditional spectral response explicit~\cite{BakrCQED2025}.

For the usual dispersive interaction, $H_{\mathrm{int}}=\hbar\chi\sigma_z\hat n$ commutes with $\sigma_z$. The qubit state therefore shifts the resonator response without, in the ideal dispersive limit, requiring a transition between $\ket g$ and $\ket e$. Photons leaking into the transmission line carry the state-dependent response outward. The resonator is the pointer, the transmission line provides the environmental channel, and the amplified microwave signal is the classical record. This separation is useful pedagogically. The recorded voltage is not the abstract qubit state itself. It is a physical record correlated with alternatives of that state.

The useful teaching point is modest but general. A symbol becomes an experimentally meaningful measurement outcome because a measurement interaction correlates alternatives of the system with distinguishable records. Its appearance as an argument of a wavefunction is not enough. In the position experiment this correlation is produced by a position-sensitive detector. In cQED it is produced by state-conditioned electromagnetic response. The Born rule assigns probabilities once the measurement structure has been specified. It does not by itself decide which observable an apparatus measures. Figure~\ref{fig:cqed} makes the distinction concrete. The qubit state belongs to the quantum description. The measured $I(t)$ and $Q(t)$ samples are classical records indexed by laboratory time. A transition during readout may therefore appear as a change in the statistics or path of the record. The record is not literally the quantum-state trajectory itself.

The distinction becomes especially transparent when the full time trace is retained rather than compressed to one integrated $I$--$Q$ point. In the path-signature readout experiment of Cao \emph{et al.}~\cite{CaoPathSignature2024}, the acquired response is treated as the classical complex-valued record
\begin{equation}
dX(t)=\{I(t),Q(t)\}\,dt,\qquad 0\le t\le T_r ,
\label{eq:cqedrecord}
\end{equation}
where $T_r$ is the externally prescribed readout duration. Conventional discrimination forms a weighted endpoint
\begin{equation}
\widetilde R=\int_0^{T_r}w(t)\,dX(t),
\label{eq:cqedendpoint}
\end{equation}
which deliberately forgets how that endpoint was reached. The signature construction instead first retains the cumulative record
\begin{equation}
X(t)=\int_0^t w(\tau)\,dX(\tau),
\label{eq:cqedpath}
\end{equation}
and evaluates ordered iterated integrals of this path. Its first level is just the total displacement $X(T_r)-X(0)$ and therefore recovers the conventional integrated readout. At second level the antisymmetric combination
\begin{equation}
A=\frac12\left[
\int_{t_1<t_2}dI(t_1)dQ(t_2)
-
\int_{t_1<t_2}dQ(t_1)dI(t_2)
\right]
\label{eq:levyarea}
\end{equation}
is the L\'evy area and is sensitive to the ordered geometry of the $I$--$Q$ path. Two records may therefore end at nearly the same integrated point while having different histories. A mid-readout transition changes the state-conditioned resonator response and can bend the measured path. The path signature can retain that information. It was used to classify such transitions and improve the inferred state at the end of the measurement.

This provides a particularly clean state-versus-history example. The object to which the signature is applied is the digitized \emph{classical measurement record}, not the abstract quantum state. A conditioned quantum trajectory such as $\rho_c(t)$ requires both the measurement record and a quantum measurement model. The path-signature work instead uses the record to infer initial and final state labels and transition classes. It does not identify $X(t)$ with $\rho_c(t)$. In one of the data sets, laboratory time is appended as a third path coordinate. This procedure is called ``time augmentation.'' It is a deliberate feature-engineering step. The external label $t$ is copied into the classical record space so that the classifier can retain information about \emph{when} a geometric feature occurs within the readout window. This does not turn $t$ into a quantum observable.

\begin{figure*}[t!]
\centering
\begin{tikzpicture}[>=Latex,font=\scriptsize,x=1cm,y=1cm]
% Panel A: physical chain
\node[anchor=west,font=\bfseries] at (-7.2,3.0) {(a) Quantum system, pointer, environment, record};
\draw[rounded corners,thick,teal] (-7.15,-0.15) rectangle (-1.0,2.65);
\node[draw,rounded corners,minimum width=1.45cm,minimum height=0.75cm,align=center] (q) at (-6.3,1.35) {transmon\\$\ket g,\ket e$};
\draw[thick] (-4.95,1.35)--(-3.05,1.35);
\draw[thick] (-4.95,1.08)--(-4.95,1.62);\draw[thick] (-3.05,1.08)--(-3.05,1.62);
\draw[decorate,decoration={snake,amplitude=1.4pt,segment length=5pt},blue] (-4.75,1.35)--(-3.25,1.35);
\node[above] at (-4.0,1.58) {resonator pointer};
\draw[<->,very thick] (q.east)--(-4.95,1.35) node[midway,above] {$\chi$};
\draw[->,very thick] (-3.05,1.35)--(-1.85,1.35) node[midway,above] {$\kappa$};
\draw[thick] (-1.85,1.35)--(-1.1,1.35);
\node[below,align=center] at (-6.3,0.72) {quantum\\system};
\node[below,align=center] at (-4.0,0.72) {state-dependent\\electromagnetic response};
\node[below,align=center] at (-1.68,0.72) {outgoing\\continuum};

\draw[->,very thick] (-0.75,1.35)--(0.25,1.35);
\node[draw,rounded corners,minimum width=1.35cm,minimum height=0.75cm,align=center] (amp) at (1.0,1.35) {amplifier\\+ digitizer};
\draw[->,very thick] (1.75,1.35)--(2.65,1.35);
\node[draw,rounded corners,minimum width=1.35cm,minimum height=0.75cm,align=center] (rec) at (3.45,1.35) {classical\\record};
\node[below,align=center] at (1.0,0.72) {measurement\\chain};
\node[below,align=center] at (3.45,0.72) {$I(t),Q(t)$ or\\assigned bit};

% Panel B: state-conditioned spectral response
\node[anchor=west,font=\bfseries\scriptsize] at (-7.2,-0.72) {(b) State-conditioned resonator response};
\draw[->] (-6.75,-3.75)--(-1.15,-3.75) node[right] {$\omega$};
\draw[->] (-6.5,-3.9)--(-6.5,-1.35) node[above] {response};
\draw[blue,very thick,domain=-6.3:-1.6,samples=120,smooth]
 plot (\x,{-3.68+1.8/(1+24*(\x+4.45)*(\x+4.45))});
\draw[orange,very thick,domain=-6.3:-1.6,samples=120,smooth]
 plot (\x,{-3.68+1.8/(1+24*(\x+3.55)*(\x+3.55))});
\draw[dashed] (-4.45,-3.75)--(-4.45,-1.7);\node[below] at (-4.45,-3.82) {$\omega_r^{g}$};
\draw[dashed] (-3.55,-3.75)--(-3.55,-1.7);\node[below] at (-3.55,-3.82) {$\omega_r^{e}$};
\node at (-5.55,-1.72) {$\ket g$ load};
\node at (-2.55,-1.72) {$\ket e$ load};
\draw[red,dashed,thick] (-4.0,-3.75)--(-4.0,-1.55);\node[right] at (-3.92,-1.35) {fixed probe};

% Panel C: two pointer records
\node[anchor=west,font=\bfseries\scriptsize] at (0.15,-0.72) {(c) Distinguishable $I$--$Q$ records};
\draw[->] (0.25,-3.75)--(5.8,-3.75) node[right] {$I$};
\draw[->] (0.55,-4.0)--(0.55,-1.35) node[above] {$Q$};
\draw[blue,thick] (1.4,-2.35) circle (0.55);\fill[blue] (1.4,-2.35) circle (2pt);\node[below] at (1.4,-2.95) {$\ket g$ pointer};
\draw[orange,thick] (4.15,-2.05) circle (0.55);\fill[orange] (4.15,-2.05) circle (2pt);\node[below] at (4.15,-2.65) {$\ket e$ pointer};
\draw[->,thick] (2.0,-2.3)--(3.5,-2.08) node[midway,above] {};
\node[align=center] at (3.0,-3.45) {};
\end{tikzpicture}
\caption{A more explicit circuit-QED measurement example. (a) The qubit changes the resonator boundary load. The resonator acts as a pointer. Photons escaping into the transmission line carry the state-dependent response, and room-temperature electronics produce a classical record. (b) In the dispersive regime the two qubit states shift the resonator response to different frequencies. A probe at one fixed frequency therefore acquires a state-dependent amplitude and phase. (c) Repeated shots produce distinguishable pointer distributions in the measured $I$--$Q$ plane. The quantum state is not directly plotted. The plotted points are classical records whose statistics depend on the state.}
\label{fig:cqed}
\end{figure*}

\subsection{An optical atomic clock with prescribed interrogation time and measured spin}
\label{subsec:atomicclock}

An optical atomic clock supplies the complementary laboratory example. In the Oxford trapped-ion experiment of Nichol \emph{et al.}~\cite{Nichol2022}, the clock transition of a trapped $^{88}\mathrm{Sr}^{+}$ ion is interrogated by a Ramsey sequence. A superposition evolves for a deliberately chosen Ramsey duration $T_R$ between two $\pi/2$ pulses. If $\Delta_i=\omega_L-\omega_i$ is the detuning between the probe laser and the $i$th atomic transition, the measured spin signal obeys
\begin{equation}
\langle \hat\Pi_i\rangle
=C_i\cos\!\left(\Delta_iT_R+\phi_i\right),
\qquad
\hat\Pi_i=\sigma_{z i}.
\label{eq:atomicclockramsey}
\end{equation}
The roles are therefore sharply separated. $T_R$ is the externally programmed duration during which the state evolves. The random quantum outcome is the final spin result. Repeated spin measurements are then used to infer the frequency detuning. In the two-clock experiment the directly measured two-ion observable is the parity $\hat\Pi=\sigma_{z1}\sigma_{z2}$. The clock--clock frequency difference is inferred from this measurement. Calling the ion a ``clock'' therefore does not mean that the Schr\"odinger parameter $t$ has become an eigenvalue of a time operator. In this experiment the atom functions as a stable frequency and phase reference. The laboratory control system supplies the Ramsey duration.

The comparison with cQED is instructive. In cQED, external time labels the samples. The apparatus records $I(t)$ and $Q(t)$ and uses them to infer the qubit state. In Ramsey clock spectroscopy, external time fixes $T_R$. The apparatus records spin or parity outcomes and uses them to infer a frequency. In both cases the experimentally useful number called a ``time'' first enters as part of the controlled protocol. The measurement interaction and detector determine what becomes a quantum measurement outcome.

\teachingbox{Put the two experiments side by side. In cQED, ask which quantities are digitizer outcomes and which quantity timestamps them. In the ion clock, ask whether the measured object is $T_R$ or $\sigma_z$. The answers are $I,Q$ versus the external label $t$, and $\sigma_z$ versus the externally chosen Ramsey duration $T_R$.}

\section{A compact teaching sequence}

The discussion can be taught without beginning with the full literature on time operators. We recommend the following progression.

\begin{enumerate}
\item \textbf{Start with the abstract state.} Write $\ket{\psi_t}=U(t)\ket{\psi_0}$ before introducing a wavefunction. Ask what changes when a basis is chosen.
\item \textbf{Introduce a spectral representation.} Write $\psi_t(x)=W_X\ket{\psi_t}$ and emphasize that $x$ enters at this second step.
\item \textbf{Delay generalized kets.} Derive the finite-grid normalization and only then take the continuum limit to motivate $\delta(x-x')$ and $\ket{x}$ notation.
\item \textbf{Separate the five roles in Table~\ref{tab:taxonomy}.} In particular, compare the spatial translation parameter with the temporal translation parameter before comparing position and time observables.
\item \textbf{Exchange prescribed and measured variables.} Contrast ``position at time $t_0$'' with ``arrival time at position $L$.''
\item \textbf{Use laboratory records.} Compare the cQED $I$--$Q$ time trace with Ramsey clock spectroscopy. In each case identify the externally prescribed time label, the physical pointer or detector record, and the quantity inferred from that record.
\item \textbf{Introduce Pauli last.} State the global assumptions that produce the spectral contradiction, identify which assumptions are relaxed in operational time measurements, and then contrast the obstruction with the exactly solvable $H=vP$ case of Sec.~\ref{subsec:canonicaltime}.
\end{enumerate}

Table~\ref{tab:misconceptions} turns this sequence into diagnostic questions that can be used in tutorials.

\begin{table*}[t]
\caption{Misconceptions and short corrections.}
\label{tab:misconceptions}
\small
\setlength{\tabcolsep}{3pt}
\begin{tabular}{p{5.1cm}p{10.25cm}}
\toprule
Statement & Diagnostic correction\\
\midrule
``$x$ and $t$ are both arguments of $\psi(x,t)$, so they have the same status.'' & $t$ first labels the evolved abstract state. The label $x$ enters after a position spectral representation is chosen.\\
``If $\psi(x)=\braket{x}{\psi}$, then $\psi(t)$ must require $\bra{t}$.'' & No. A time bra is needed only if a particular time-valued observable/representation has been defined. Ordinary Schr\"odinger $t$ labels evolution.\\
``Position is the spatial analogue of the Stone time parameter.'' & The Stone pairings are displacement--momentum and time displacement--Hamiltonian. Position is a covariant observable under spatial translations.\\
``A state has a definite time but an indefinite position.'' & The external $t$ labels which state is being discussed. A quantum clock included inside the model can itself have an uncertain reading.\\
``Pauli proved that time cannot be observable.'' & The strong obstruction concerns a globally canonical self-adjoint $T$ together with a semibounded $H$. Restricted-domain operators and time POVMs are not excluded by that statement.\\
\bottomrule
\end{tabular}
\end{table*}

\section{Discussion of the real asymmetry}

After the distinctions above, there is no single yes-or-no answer to ``Are space and time asymmetric in quantum mechanics?'' The answer depends on what is being compared.

If one compares external background coordinates, both $x$ and $t$ are parameters in the description. If one compares dynamical quantities carried by physical systems, both particle position and clock readings can be quantum observables. If one compares the standard nonrelativistic formulation of a subsystem, however, the theory normally takes an external time parameter as the label of dynamical evolution while treating the subsystem's position as an observable. That is a modelling and kinematical distinction built into the usual formulation, not a conclusion derived solely from Pauli's theorem.

The derivative notation can therefore also mislead. In the Schr\"odinger equation, $\partial_t$ differentiates the family $t\mapsto\ket{\psi_t}$ after a representation has been chosen. In the position representation, $\partial_x$ acts within a given function over the representation coordinate $x$. At fixed $t$, changing $x$ does not move between different collapsed outcome states. It compares neighboring amplitudes within the representation of the same quantum state. Thus $\partial_t$ and $\partial_x$ may appear next to one another in a partial differential equation while differentiating along different conceptual directions. The first differentiates along the dynamical family of states. The second differentiates within the coordinate representation of each state.

This contrast can also be discussed analytically. Callender has argued that a deeper temporal asymmetry can be sought in the directions along which physical laws furnish well-posed Cauchy problems~\cite{Callender2017}. In a well-posed initial-value problem, appropriate initial data determine a unique solution that depends continuously on those data. His ``sideways'' analysis asks what happens when physical equations are instead treated as evolution equations in non-temporal directions and uses the resulting well-posedness question to sharpen the distinction. On this view, the difference between $\partial_t$ and $\partial_x$ is not only a difference in what each ranges over. Well-posedness supplies an additional criterion for identifying the temporal direction. Our taxonomy is compatible with this proposal but does not presuppose it.

This distinction does not deny that spacetime symmetries, relativistic covariance, constrained systems, or relational clock constructions can reorganize these roles. It simply prevents notation from deciding the ontology in advance.

The taxonomy also connects the classroom discussion to philosophy-of-physics literature that instructors can point students toward. Hilgevoord and Uffink's Stanford Encyclopedia entry surveys the historical and conceptual difficulties surrounding uncertainty relations, including the special status of time in time--energy relations~\cite{SEPUncertainty}. Butterfield's survey of time in quantum physics organizes the subject around the same external, intrinsic, and observable trichotomy used here~\cite{Butterfield2013}. It also explains which interpretive questions remain open. The ``problem of time'' in canonical quantum gravity arises when the usual external-time option is unavailable; the Hamiltonian constraint leaves physical states without evolution relative to a background $t$ in the ordinary Schr\"odinger sense~\cite{SEPQuantumGravity}. Relational clock constructions such as Page--Wootters provide one possible response~\cite{PageWootters1983,GLM2015}. Table~\ref{tab:taxonomy} can be read as the nonrelativistic, unconstrained corner of that larger discussion.

\section{Conclusion}

The question ``Does $\psi(t)$ require a $\bra{t}$?'' is useful precisely because the immediate answer, ``no,'' is less important than understanding why. The notation $\psi(x,t)$ hides a composition of operations. The external evolution parameter selects a state along a dynamical trajectory. The spectral representation of an observable then turns that state into an amplitude function labelled by possible outcomes. Generalized position kets are optional shorthand for that spectral representation, and their delta normalization can be motivated transparently from an ordinary finite grid. A time-valued physical record can become an observable, but that is a different construction from the external parameter used to evolve the state. Pauli's argument then constrains a strong global canonical time operator rather than forbidding every possible time measurement. The $H=vP$ boundary case makes the limitation explicit: once the relevant spectrum is two-sided, a self-adjoint observable $\Theta=X/v$ can satisfy the global canonical covariance relation while the Schr\"odinger $t$ remains the external evolution parameter.

For students and instructors, the practical rule is simple. Whenever a symbol appears in a quantum formula, ask three questions before assigning it physical meaning.
\begin{enumerate}
\item Does it label the \emph{state's evolution}?
\item Does it label a \emph{spectral representation or possible outcome}?
\item What \emph{measurement interaction and record} would make that label operational?
\end{enumerate}
The same letter may answer different questions in different experiments. Same notation does not imply same mathematical role.

\appendix
\section{Classroom problems with brief instructor solutions}

\subsection{Problem 1. A time-dependent spin without $\ket{t}$}
A spin-$1/2$ particle in a static magnetic field has
\begin{equation}
\ket{\psi(t)}=c_+(t)\ket{+z}+c_-(t)\ket{-z}.
\end{equation}
Ask students three questions. What is the representation label? What labels evolution? Is $\bra{t}$ needed?

\emph{Instructor solution.} The spectral labels are $\pm1$ (or $\pm\hbar/2$) for the chosen spin observable. The parameter $t$ labels the family of states. No time bra is implied. The example isolates the distinction from all continuous-spectrum subtleties.

\subsection{Problem 2. Where did the delta function come from?}
Start with $N$ orthonormal grid states of spacing $\Delta x$. Ask students to rescale the states so that the probability in one cell is approximately $|\psi(x_j)|^2\Delta x$ and derive the resulting inner product and completeness relation.

\emph{Instructor solution.} The rescaling $\ket{x_j}_c=\ket{j}/\sqrt{\Delta x}$ gives ${}_c\braket{x_j}{x_k}_c=\delta_{jk}/\Delta x$ and $\I=\sum_j\Delta x\ket{x_j}_c{}_c\bra{x_j}$. The continuum limit motivates both $\delta(x-x')$ and $\int dx\ket{x}\bra{x}=\I$.

\subsection{Problem 3. Translate versus measure}
A wavepacket is translated to the right by $a$. Ask students to identify the group parameter, its generator, the position observable, and the possible position outcome.

\emph{Instructor solution.} The group parameter is $a$, the generator is $P$, the observable is $X$, and the random measurement outcome is a numerical position $x$. Their relation is covariance, not identity.

\subsection{Problem 4. Two detector protocols}
For a wavepacket incident on a detector at $L$, compare (A) measuring position at a prescribed $t_0$ and (B) recording the first detection time at prescribed $L$. Is $|\psi(L,t)|^2dt$ automatically the probability of a click in $[t,t+dt]$?

\emph{Instructor solution.} No. Protocol A is a position measurement conditioned on externally selected time. Protocol B asks for a time-of-event observable/instrument. Its distribution depends on the measurement model or an appropriate arrival-time POVM and is not generally obtained by renormalizing $|\psi(L,t)|^2$.

\subsection{Problem 5. Locate the assumption in Pauli's argument}
Suppose a proposed $T$ satisfies $[T,H]=i\hbar$ on some dense domain. Can one immediately conclude that $e^{isT/\hbar}$ shifts every energy by $s$ for every real $s$?

\emph{Instructor solution.} No. Exponentiating an unbounded-operator commutator requires domain and global covariance assumptions. Pauli's spectral contradiction follows from the stronger global translation property, not from every formal restricted-domain commutator.


\begin{thebibliography}{99}

\bibitem{Pauli1933}
W.~Pauli,
``Die allgemeinen Prinzipien der Wellenmechanik,''
in \emph{Handbuch der Physik}, 2nd ed., edited by H.~Geiger and K.~Scheel, Vol.~24, Part~1
(Springer, Berlin, 1933), pp.~83--272.
English translation in \textit{General Principles of Quantum Mechanics}
(Springer, Berlin, 1980), see the footnote on p.~63.

\bibitem{MarshmanSingh2015}
E.~Marshman and C.~Singh,
``Student difficulties with quantum states while translating state vectors in Dirac notation to wave functions in position and momentum representations,''
\href{https://doi.org/10.1119/perc.2015.pr.048}
{in \emph{Physics Education Research Conference 2015}, 211--214 (2015)}.

\bibitem{Emigh2015}
P.~J.~Emigh, G.~Passante, and P.~S.~Shaffer,
``Student understanding of time dependence in quantum mechanics,''
\href{https://doi.org/10.1103/PhysRevSTPER.11.020112}
{\emph{Phys. Rev. ST Phys. Educ. Res.} \textbf{11}, 020112 (2015)}.

\bibitem{delaTorre2002}
A.~C.~de~la~Torre,
``Relativity of representations in quantum mechanics,''
\href{https://doi.org/10.1119/1.1419100}
{\emph{Am. J. Phys.} \textbf{70}, 298--300 (2002)}.

\bibitem{Styer2002}
D.~F.~Styer, M.~S.~Balkin, K.~M.~Becker \emph{et al.},
``Nine formulations of quantum mechanics,''
\href{https://doi.org/10.1119/1.1445404}
{\emph{Am. J. Phys.} \textbf{70}, 288--297 (2002)}.

\bibitem{Hilgevoord2002}
J.~Hilgevoord,
``Time in quantum mechanics,''
\href{https://doi.org/10.1119/1.1430697}
{\emph{Am. J. Phys.} \textbf{70}, 301--306 (2002)}.

\bibitem{Hilgevoord2005}
J.~Hilgevoord,
``Time in quantum mechanics: A story of confusion,''
\href{https://doi.org/10.1016/j.shpsb.2004.10.002}
{\emph{Stud. Hist. Philos. Mod. Phys.} \textbf{36}, 29--60 (2005)}.

\bibitem{Busch1990}
P.~Busch,
``On the energy--time uncertainty relation. Part I: Dynamical time and time indeterminacy,''
\href{https://doi.org/10.1007/BF00732932}
{\emph{Found. Phys.} \textbf{20}, 1--32 (1990)}.
``On the energy--time uncertainty relation. Part II: Pragmatic time versus energy indeterminacy,''
\href{https://doi.org/10.1007/BF00732933}
{\emph{Found. Phys.} \textbf{20}, 33--43 (1990)}.

\bibitem{Busch2008}
P.~Busch,
``The time--energy uncertainty relation,''
\href{https://doi.org/10.1007/978-3-540-73473-4_3}
{in \emph{Time in Quantum Mechanics}, 2nd ed., edited by J.~G.~Muga, R.~Sala~Mayato, and I.~L.~Egusquiza,
Lecture Notes in Physics Vol.~734 (Springer, Berlin Heidelberg, 2008), pp.~73--105}.

\bibitem{Peres1980}
A.~Peres,
``Measurement of time by quantum clocks,''
\href{https://doi.org/10.1119/1.12061}
{\emph{Am. J. Phys.} \textbf{48}, 552--557 (1980)}.

\bibitem{MugaLeavens2000}
J.~G.~Muga and C.~R.~Leavens,
``Arrival time in quantum mechanics,''
\href{https://doi.org/10.1016/S0370-1573(00)00047-8}
{\emph{Phys. Rep.} \textbf{338}, 353--438 (2000)}.

\bibitem{MugaBook2008}
J.~G.~Muga, R.~Sala~Mayato, and I.~L.~Egusquiza, eds.,
\textit{Time in Quantum Mechanics}, 2nd ed., Lecture Notes in Physics Vol.~734
\href{https://doi.org/10.1007/978-3-540-73473-4}
{(Springer, Berlin Heidelberg, 2008)}.

\bibitem{Roberts1966}
J.~E.~Roberts,
``The Dirac bra and ket formalism,''
\href{https://doi.org/10.1063/1.1705001}
{\emph{J. Math. Phys.} \textbf{7}, 1097--1104 (1966)}.

\bibitem{Gieres2000}
F.~Gieres,
``Mathematical surprises and Dirac's formalism in quantum mechanics,''
\href{https://doi.org/10.1088/0034-4885/63/12/201}
{\emph{Rep. Prog. Phys.} \textbf{63}, 1893--1931 (2000)}.

\bibitem{delaMadrid2005}
R.~de~la~Madrid,
``The role of the rigged Hilbert space in quantum mechanics,''
\href{https://doi.org/10.1088/0143-0807/26/2/008}
{\emph{Eur. J. Phys.} \textbf{26}, 287--312 (2005)}.

\bibitem{Shankar1994}
R.~Shankar,
\textit{Principles of Quantum Mechanics}, 2nd ed.
\href{https://doi.org/10.1007/978-1-4757-0576-8}
{(Plenum Press, New York, 1994)}, Chap.~1.

\bibitem{Belot2007}
G.~Belot,
``The representation of time and change in mechanics,''
\href{https://doi.org/10.1016/B978-044451560-5/50005-1}
{in \emph{Philosophy of Physics}, edited by J.~Butterfield and J.~Earman
(Elsevier, Amsterdam, 2007), pp.~133--227}.

\bibitem{ChalkerLukas2010}
J.~T.~Chalker and A.~Lukas,
\textit{M.Phys Option in Theoretical Physics: C6, Lecture Notes}
\href{https://www-thphys.physics.ox.ac.uk/people/JohnChalker/theory/lecture-notes.pdf}
{(University of Oxford, Oxford, 2010--2011), Chap.~1, Sec.~1.1.3}.

\bibitem{Zee2010}
A.~Zee,
\textit{Quantum Field Theory in a Nutshell}, 2nd ed.
(Princeton University Press, Princeton, NJ, 2010), Chap.~I.2.

\bibitem{NewtonWigner1949}
T.~D.~Newton and E.~P.~Wigner,
``Localized states for elementary systems,''
\href{https://doi.org/10.1103/RevModPhys.21.400}
{\emph{Rev. Mod. Phys.} \textbf{21}, 400--406 (1949)}.

\bibitem{Hegerfeldt1974}
G.~C.~Hegerfeldt,
``Remark on causality and particle localization,''
\href{https://doi.org/10.1103/PhysRevD.10.3320}
{\emph{Phys. Rev. D} \textbf{10}, 3320--3321 (1974)}.

\bibitem{Malament1996}
D.~B.~Malament,
``In defense of dogma: Why there cannot be a relativistic quantum mechanics of (localizable) particles,''
\href{https://doi.org/10.1007/978-94-015-8656-6_1}
{in \emph{Perspectives on Quantum Reality}, edited by R.~Clifton
(Kluwer, Dordrecht, 1996), pp.~1--10}.

\bibitem{Allcock1969}
G.~R.~Allcock,
``The time of arrival in quantum mechanics I. Formal considerations,''
\href{https://doi.org/10.1016/0003-4916(69)90251-6}
{\emph{Ann. Phys.} \textbf{53}, 253--285 (1969)}.
See also ``II. The individual measurement,''
\href{https://doi.org/10.1016/0003-4916(69)90252-8}
{\emph{Ann. Phys.} \textbf{53}, 286--310 (1969)}, and
``III. The measurement ensemble,''
\href{https://doi.org/10.1016/0003-4916(69)90253-X}
{\emph{Ann. Phys.} \textbf{53}, 311--348 (1969)}.

\bibitem{Kijowski1974}
J.~Kijowski,
``On the time operator in quantum mechanics and the Heisenberg uncertainty relation for energy and time,''
\href{https://doi.org/10.1016/S0034-4877(74)80004-2}
{\emph{Rep. Math. Phys.} \textbf{6}, 361--386 (1974)}.

\bibitem{DasDurr2019}
S.~Das and D.~D\"urr,
``Arrival time distributions of spin-$1/2$ particles,''
\href{https://doi.org/10.1038/s41598-018-38261-4}
{\emph{Sci. Rep.} \textbf{9}, 2242 (2019)}.

\bibitem{GoldsteinTumulkaZanghi2024}
S.~Goldstein, R.~Tumulka, and N.~Zangh\`{\i},
``On the spin dependence of detection times and the nonmeasurability of arrival times,''
\href{https://doi.org/10.1038/s41598-024-53777-8}
{\emph{Sci. Rep.} \textbf{14}, 3775 (2024)}.

\bibitem{Galapon2002}
E.~A.~Galapon,
``Pauli's theorem and quantum canonical pairs: The consistency of a bounded, self-adjoint time operator canonically conjugate to a Hamiltonian with non-empty point spectrum,''
\href{https://doi.org/10.1098/rspa.2001.0874}
{\emph{Proc. R. Soc. Lond. A} \textbf{458}, 451--472 (2002)}.

\bibitem{PageWootters1983}
D.~N.~Page and W.~K.~Wootters,
``Evolution without evolution: Dynamics described by stationary observables,''
\href{https://doi.org/10.1103/PhysRevD.27.2885}
{\emph{Phys. Rev. D} \textbf{27}, 2885--2892 (1983)}.

\bibitem{GLM2015}
V.~Giovannetti, S.~Lloyd, and L.~Maccone,
``Quantum time,''
\href{https://doi.org/10.1103/PhysRevD.92.045033}
{\emph{Phys. Rev. D} \textbf{92}, 045033 (2015)}.

\bibitem{Holevo1982}
A.~S.~Holevo,
\textit{Probabilistic and Statistical Aspects of Quantum Theory}
(North-Holland, Amsterdam, 1982).

\bibitem{Blais2004}
A.~Blais, R.-S.~Huang, A.~Wallraff, S.~M.~Girvin, and R.~J.~Schoelkopf,
``Cavity quantum electrodynamics for superconducting electrical circuits: An architecture for quantum computation,''
\href{https://doi.org/10.1103/PhysRevA.69.062320}
{\emph{Phys. Rev. A} \textbf{69}, 062320 (2004)}.

\bibitem{Blais2021}
A.~Blais, A.~L.~Grimsmo, S.~M.~Girvin, and A.~Wallraff,
``Circuit quantum electrodynamics,''
\href{https://doi.org/10.1103/RevModPhys.93.025005}
{\emph{Rev. Mod. Phys.} \textbf{93}, 025005 (2021)}.

\bibitem{BakrCQED2025}
M.~Bakr,
``A Boundary Condition Perspective on Circuit QED Dispersive Readout,''
\href{https://arxiv.org/abs/2512.24466}
{arXiv:2512.24466 [quant-ph] (2025)}.

\bibitem{CaoPathSignature2024}
S.~Cao, Z.~Shao, J.-Q.~Zheng \emph{et al.},
``Superconducting qubit readout enhanced by path signature,''
\href{https://arxiv.org/abs/2402.09532}
{arXiv:2402.09532 [quant-ph] (2024)}.

\bibitem{Nichol2022}
B.~C.~Nichol, R.~Srinivas, D.~P.~Nadlinger \emph{et al.},
``An elementary quantum network of entangled optical atomic clocks,''
\href{https://doi.org/10.1038/s41586-022-05088-z}
{\emph{Nature} \textbf{609}, 689--694 (2022)}.

\bibitem{Callender2017}
C.~Callender,
``Looking at the world sideways,''
\href{https://doi.org/10.1093/oso/9780198797302.003.0008}
{in \textit{What Makes Time Special?}
(Oxford University Press, Oxford, 2017), Chap.~8, pp.~157--179}.

\bibitem{SEPUncertainty}
J.~Hilgevoord and J.~Uffink,
``The uncertainty principle,''
\href{https://plato.stanford.edu/archives/spr2024/entries/qt-uncertainty/}
{in \emph{The Stanford Encyclopedia of Philosophy}, Spring 2024 ed., edited by E.~N.~Zalta and U.~Nodelman
(Metaphysics Research Lab, Stanford University, 2024)}.

\bibitem{Butterfield2013}
J.~Butterfield,
``On time in quantum physics,''
\href{https://doi.org/10.1002/9781118522097.ch14}
{in \emph{A Companion to the Philosophy of Time}, edited by H.~Dyke and A.~Bardon
(Wiley-Blackwell, Chichester, 2013), pp.~220--241}.

\bibitem{SEPQuantumGravity}
S.~Weinstein and D.~Rickles,
``Quantum gravity,''
\href{https://plato.stanford.edu/archives/spr2024/entries/quantum-gravity/}
{in \emph{The Stanford Encyclopedia of Philosophy}, Spring 2024 ed., edited by E.~N.~Zalta and U.~Nodelman
(Metaphysics Research Lab, Stanford University, 2024)}.

\end{thebibliography}
\end{document}